\documentclass[lettersize,journal]{IEEEtran}

\usepackage[dvips]{graphicx}
\usepackage{latexsym}
\usepackage{epsfig}
\usepackage{color}
\usepackage{amsmath} % real numbers, natural numbers, ...
\usepackage{amssymb}
\usepackage{amsxtra}
\usepackage{enumitem}
\usepackage{float}
\usepackage[T1]{fontenc}% optional T1 font encoding
\usepackage{setspace}
\usepackage{balance}
\usepackage{diagbox}
\usepackage{epstopdf}
\usepackage{amsthm}
\usepackage{dsfont} % for \mathds{N}
\usepackage{multirow}
\usepackage{hyperref}
\usepackage{tabularx,colortbl}
\usepackage[table]{xcolor}
\usepackage{lscape}
\usepackage{url}
\usepackage{cite}
\usepackage{algorithm}
\usepackage{algpseudocode}
\usepackage[ subrefformat=parens,labelformat=parens, caption=false, font=footnotesize]{subfig}
\usepackage{mathtools}
\usepackage{units}

\usepackage{flushend}

\usepackage{cases}

\newtheorem{proposition}{\textbf{Proposition}}

\newtheorem{remark}{\textbf{Remark}}

\usepackage[cmintegrals]{newtxmath}

\makeatletter
\def\BState{\State\hskip-\ALG@thistlm}
\makeatother

\mathchardef\mhyphen="2D

\newcommand{\nth}[1]{{#1}{\text{th}}}

\newcommand{\abs}[1]{\left|{#1}\right|}

\newtheorem{theorem}{Theorem}

\makeatother

\makeatletter
\def\ps@IEEEtitlepagestyle{%
  \def\@oddfoot{\mycopyrightnotice}%
  \def\@oddhead{\hbox{}\@IEEEheaderstyle\leftmark\hfil\thepage}\relax
  \def\@evenhead{\@IEEEheaderstyle\thepage\hfil\leftmark\hbox{}}\relax
  \def\@evenfoot{}%
}

\def\mycopyrightnotice{%
  \begin{minipage}{\textwidth}
  \centering \scriptsize
    This work has been accepted by the IEEE Communications Letters for publication.  Copyright may be transferred without notice, after which this version may no longer be accessible.
  \end{minipage}
}
\makeatother

\begin{document}

\title{Performance Analysis of Outdoor THz Links under Mixture Gamma Fading with Misalignment}

\author{Hakim~Jemaa,~\IEEEmembership{Student Member,~IEEE,}
        Simon Tarboush,
        Hadi~Sarieddeen,~\IEEEmembership{Member,~IEEE,}
        
        Mohamed-Slim~Alouini,~\IEEEmembership{Fellow,~IEEE,}
        and~Tareq~Y.~Al-Naffouri,~\IEEEmembership{Senior Member,~IEEE}% <-this % stops a space
\thanks{The work was supported by the King Abdullah University of Science and Technology (KAUST) Office of Sponsored Research under Award ORA-CRG2021-4695, and the American University of Beirut (AUB) University Research Board. Hakim~Jemaa, Mohamed-Slim~Alouini, and Tareq~Y.~Al-Naffouri are with the Department of Computer, Electrical and Mathematical Sciences and Engineering (CEMSE), KAUST, Kingdom of Saudi Arabia (\{hakim.jemaa, slim.alouini, tareq.alnaffouri\}@kaust.edu.sa).
S. Tarboush is a researcher from Damascus, Syria (simon.w.tarboush@gmail.com). H. Sarieddeen is with the ECE Department, AUB, Lebanon (hs139@aub.edu.lb).}% <-this % stops a space
%\thanks{Manuscript received xxx, xxx; revised xxx, xxx.}
}

\maketitle

\begin{abstract}
The precision of link-level theoretical performance analysis for emerging wireless communication paradigms is critical. Recent studies have demonstrated the excellent fitting capabilities of the mixture gamma (MG) distribution in representing small-scale fading in outdoor terahertz (THz)-band scenarios. Our study establishes an in-depth performance analysis for outdoor point-to-point THz links under realistic configurations, incorporating MG small-scale fading combined with the misalignment effect. We derive closed-form expressions for the bit-error probability, outage probability, and ergodic capacity. Furthermore, we conduct an asymptotic analysis of these metrics at high signal-to-noise ratios and derive the necessary convergence conditions. Simulation results, leveraging precise measurement-based channel parameters in various configurations, closely align with the derived analytical equations.

\end{abstract}


\maketitle
\section{Introduction}
\IEEEPARstart{T}{he} terahertz (THz) band, spanning from 100 gigahertz (GHz) to 10 THz, is set to revolutionize future wireless communications, potentially enabling terabit-per-second data rates and ultra-low latency~\cite{sarieddeen2020overview,Jornet2024Evolution}. THz channels exhibit extreme sparsity in time and spatial domains~\cite{sheikh2022thz,tarboush9591285}, allowing for feasible communication even via non-line-of-sight (non-LoS) paths, as sent measurements~\cite{sheikh2022thz}. In indoor environments, the number of non-LoS paths is generally small and further decreases with the use of high-gain antennas~\cite{tarboush9591285}. 
%However, precise channel modeling across scenarios requires additional measurements and validation~\cite{Jornet2024Evolution}.

Recent empirical studies have identified the Gaussian mixture (GM) and mixture gamma (MG) distributions as effective models for small-scale fading in outdoor THz links \cite{papasotiriou2023outdoor,karakoca2023measurement}. While highly-correlated THz channels pose challenges for conventional Nakagami-m and Rician distributions \cite{papasotiriou2021experimentally}, measurements have verified the adaptability of MG and GM distributions, particularly in multiple-peak scenarios caused by multipath effects at sub-THz frequencies \cite{papasotiriou2023outdoor}. The gamma distribution's inherent non-negative nature, robustness to outliers in high-correlation scenarios, and ability to handle multiple peaks favor the MG distribution \cite{papasotiriou2023outdoor,karakoca2023measurement}. Early work on MG fading \cite{atapattu2011mixture} in typical wireless channels explored its applicability for signal-to-noise ratio (SNR) modeling and symbol error probability without focusing on THz channels; the MG distribution was used to study the performance of diversity reception schemes over generalized-K fading channels in \cite{jung2014capacity}.

The $\alpha$-$\mu$ distribution accurately models indoor THz small-scale fading with versatile parameterization \cite{papasotiriou2021experimentally}. This distribution is leveraged to construct an analytical framework for indoor THz channels in \cite{8610080}, accounting for the joint effect of misalignment fading using the zero-boresight model; a different pointing error distribution is used in \cite{10018285}. The work in \cite{9714471} further accounts for random fog modeling using a gamma distribution. For outdoor THz systems, \cite{bhardwaj2022performance} develops an analytical framework with closed-form expressions using the GM distribution for small-scale fading. In \cite{Varotsos2023capacity}, atmospheric channel effects are analyzed using a gamma-gamma distribution for outdoor THz environments. However, many previous studies neglect the necessary conditions to ensure the validity of derived expressions, which depend on realistic system parameters, making results difficult to reproduce under measurement-based THz configurations. Despite the attractive properties of the MG distribution, it has received less attention for accurately describing outdoor THz links. For example, \cite{10458985} utilizes the MG distribution to model the cascaded channel of intelligent reflected surface (IRS)-assisted communication.

In this work, we present an analytical framework for point-to-point single-input single-output (SISO) outdoor THz channels by adopting the MG distribution for small-scale fading and the zero-boresight model for misalignment fading. We study the bit-error probability, outage probability, and ergodic capacity, deriving novel closed-form expressions and ensuring necessary conditions to validate the equations. The results are expressed using the univariate Fox-H function. Additionally, we analyze the asymptotic behavior of the studied performance metrics to understand the relationship between system parameters and overall system performance. Finally, we validate our analytical study through numerical simulations using verified measurement-based parameters from the literature.

Throughout the paper, the $\abs{\cdot}$ operator represents the absolute value, $\mathbb{E}[\cdot]$ is the expectation operator, and $\mathrm{Pr}(\cdot)$ is the probability operator. $\Gamma(.)$ is the gamma function and $\Gamma(.,.)$ is the incomplete gamma function defined as: $\Gamma(\zeta, u)=\int_0^u t^{\zeta-1} e^{-t} d t$. $Q(x)=\frac{1}{\sqrt{2 \pi}} \int_x^{\infty} e^{ \left(-\frac{u^2}{2}\right)} d u$ is the $Q\mhyphen$function and $erf(x)=\frac{2}{\sqrt{\pi}} \int_0^{x} e^ {\left(-u^2\right)} d u$ is the error function \cite{prudnikov1986integrals}.$G_{p, q}^{m, n}\left[z \left\lvert \begin{smallmatrix}a_1, \ldots, a_p \\ b_1, \ldots, b_q\end{smallmatrix} \right.\right]$ represents the Meijer G-function \cite[eq. (9.301)]{prudnikov1986integrals} while $H_{p, q}^{m, n}\left[z \left\lvert \begin{smallmatrix}(a_1, b_1), \ldots, (a_p, b_p) \\ (c_1, d_1), \ldots, (c_p, d_p)\end{smallmatrix} \right.\right]$ represents the Fox H-function \cite[eq. (1.1.1)]{kilbas2004h}.

\section{System and Channel Models}
\label{sec:sysmodel}
We consider a single-carrier SISO THz link~\cite{tarboush2022single} of complex baseband equivalent received signal
\begin{equation}
\label{eq:sys_model}
    {y} = {\sqrt{P_t} h_l h_m h_f x}  +  {n}  = {\sqrt{P_t}h x}+{n},
\end{equation}
where $y$ is the received symbol, $x$ is the modulated transmitted symbol, $n$ is the additive white Gaussian noise of power $N_{0}/2$ ($N_0$ is the one-sided noise power spectral density), $P_t$ is the transmitter power, $h = h_l h_m h_f$ is the THz channel \cite{8610080}, and $h_l$, $h_m$, and $h_f$ are the free-space path loss, misalignment, and small-scale fading, respectively. We define the instantaneous SNR at the receiver, relative to the channel power, as
\begin{equation}
\gamma=E_{\mathrm{s}} P_t |h|^2/ N_{0}=\Upsilon|h|^2,
\end{equation}
where $\Upsilon=\frac{E_{\mathrm{s}}P_t}{ N_{0}}$ and $E_{\mathrm{s}}$ is the energy per symbol. 
%The next step is to describe the path loss, misalignment, and small-scale fading models used in this work. 

%\subsection{Path Loss at THz band}
THz-band path loss consists of both spreading and molecular absorption losses; it is expressed as~\cite{tarboush9591285}
\begin{equation}
    h_l= \left(c \sqrt{G_t G_r}/\left(4 \pi f d\right)\right) e^ {\left(-\frac{1}{2} \mathcal{K}_{\mathrm{abs}} d\right)},
\end{equation}
where $c$ is the speed of light, $f$ is the operating frequency, $d$ is the communication distance, $G_t$/$G_r$ are the transmit/receive antenna gains, and $\mathcal{K}_{\mathrm{abs}}$ is the molecular absorption coefficient (more details in\cite{tarboush9591285}). In measurement-based sub-THz/THz works~\cite{Jornet2024Evolution,tarboush9591285}, the path loss exponent is best-fit to 2.

%\subsection{Misalignment Fading}
We adopt the zero-boresight model to describe misalignment fading, $h_m$; its probability distribution function (PDF) is\cite{8610080}
\begin{equation}
    f_{h_{m}}(x) = \left(\rho^2/S_0^{\rho^2}\right) x^{\rho^2-1}, \quad 0 \leq x \leq S_0,
\end{equation}
where $S_0 \!=\! \mathrm{erf}(v)^2$, $v \!=\! \sqrt{\frac{\pi}{2}} \left(\frac{r_1}{\omega_z}\right)$, $ \omega_z$ is the signal beam width, $r_1$ is the effective receiver radius, $ \rho \!=\! \frac{\omega_{zeq}}{2\sigma_s} $, $\omega_{zeq} $ is the equivalent beam width at the receiving antenna, $\omega_{zeq}^2 \!=\! \frac{\omega_z^2 \sqrt{\pi} \mathrm{erf}(v)}{2v \exp(-v^2)}$, and $\sigma_s^2$ is the variance of the pointing error displacement~\cite{8610080}.

%\subsection{Outdoor THz Small-Scale Fading}
The MG distribution is used to describe the magnitude of outdoor THz small-scale fading ~\cite{papasotiriou2023outdoor,karakoca2023measurement}, and its PDF is~\cite{papasotiriou2023outdoor}
\begin{equation}
\label{MGdist}
    f_{h_f}(x)\!=\!\sum_{i=1}^K \!w_i \frac{\zeta_i^{\beta_i} x^{\beta_i-1} e^{-\zeta_i x}}{\Gamma\left(\beta_i\right)}\!=\!\sum_{i=1}^K\! \alpha _i x^{\beta _i -1} e^{-\zeta _i x}, \ x\ge 0,
\end{equation}
where $K$ is the number of gamma components, and $\zeta _i$, $\beta _i$ and $w_i$ denote the scale, shape, and weight of the $\nth{i}$ component; $\sum_{i=1}^K w_i \!=\! 1$. We define $\alpha_i\!=\!w_i\zeta_i^{\beta_i}/\Gamma\left(\beta_i\right)$ for ease of notation.

\section{Performance analysis}
\label{sec:perf_ana} 

Our objective is to construct an analytical framework for outdoor THz communications by modeling the channel based on measurement-based distributions and parameters and subsequently deriving closed-form expressions for the bit error probability, $\mathrm{Pr}_{\mathrm{e}}$, outage probability, $\mathrm{Pr}_{\mathrm{out}}$, and ergodic capacity, $C_{\mathrm{erg}}$. For a maximum likelihood detector~\cite{simon2001digital},
\begin{equation}
    \label{eq:prob_error}
    \mathrm{Pr}_{\mathrm{e}}={\mathbb{E}}_{|h|^2}\left[Q\left(a \sqrt{P_t |h|^2}\right)\right] =\int_0^{\infty} Q(a \sqrt{x}) f_{P_t |h|^2}(x) d x,
\end{equation}
where $a$ is represents a modulation-type scale (e.g, $a\!=\!2E_s/N_0$ for binary phase-shift keying (BPSK)), and $f_{|h|^2}$ is the PDF of $|h|^2$. The outage probability is expressed as~\cite{simon2001digital}
\begin{equation}
\label{eq:out}
\mathrm{Pr}_{\text {out }}=\int_0^{\gamma_{\text {thr }}} f_\gamma(\gamma) d \gamma,
\end{equation}
where $\gamma_{\text {thr }}$ is the SNR value corresponding to the information rate threshold. Finally, the ergodic capacity is defined as~\cite{8610080}
\begin{equation}
    \label{eq:capacity}
    \begin{aligned}
    C_{\mathrm{erg}} & =  \mathbb{E}_{\gamma}\left[ \log_2(1 + \gamma) \right] = \mathbb{E}_{\mathrm{|h|^2}}\left[ \log_2(1 + \Upsilon|h|^2) \right]  \\
    & = \frac{1}{\ln(2)} \int_0^{\infty} \ln (1+\Upsilon x^2) f_{|h|}(x) \mathrm{d}x.
    \end{aligned}
\end{equation}
We first obtain the distributions of $|h|$ and $P_t |h|^2$; proofs of the theorems are provided in the corresponding appendices.
\begin{theorem}[Appendix~\ref{sec:appendix_pdf}]
\label{thm:pdfh}
The PDFs of $|h|$ and $P_t |h|^2$ are
\begin{equation}
\label{thm:pdf_h}
\begin{aligned}
    & f_{|h|}(x)\!=\!\rho^2\! \sum_{i=1}^{K}\!\frac{\alpha_i x^{\beta_i-1}}{S_0^{\beta_i } |h_l|^{\beta_i}} G_{1,2}^{2,0} \left( \left. \frac{\zeta_i x}{S_0 |h_l|} \, \right| 
    \begin{array}{c}
    1 + \rho^2 - \beta_i \\
    \rho^2 - \beta_i , 0
    \end{array}\!\!\right),
\end{aligned}
\end{equation}
\begin{equation}
\label{thm:pdf_h_squared_Pt}
\begin{aligned}
    &f_{P_t |h|^2}(x) \!= \!\!\frac{\rho^2}{2} \!\!\sum_{i=1}^{K}\!C_i x^{\frac{\beta_i}{2}\!-\!1} G_{1,2}^{2,0}\!\left(\!\left. \frac{\zeta_i \sqrt{x}}{\sqrt{P_t} S_0 |h_l|} \!\right|\!\!\!\!
    \begin{array}{c}
    1\!+\!\rho^2\!-\!\beta_i\\
    \rho^2-\!\beta_i,\!0
    \end{array}\!\!\right)\!,
\end{aligned}
\end{equation}
where $C_i = \frac{\alpha_i}{P_t^{\frac{\beta_i}{2}} S_0^{\beta_i} |h_l|^{\beta_i}}$. 
\end{theorem}

\begin{remark}    
In proving~\eqref{thm:pdf_h} and~\eqref{thm:pdf_h_squared_Pt}, the Euler property of the Meijer G-function imposes certain conditions, \cite[eq. (7.811.3)]{prudnikov1986integrals}, such as $\beta_i - \rho^2 > 0, \forall i$. Our system parameters are chosen to satisfy these conditions.
\end{remark}
\subsection{Probability of error}
Using the derived PDFs of $|h|$ and $P_t |h|^2$, we derive the following theorem and proposition for $\mathrm{Pr}_{\mathrm{e}}$, with and without accounting for misalignment, respectively.
\begin{theorem}[Appendix~\ref{sec:appendix_Pe}]
\label{thm:Pe}
$\mathrm{Pr}_{\mathrm{e}}$ can be written in a closed form as
\begin{equation}
    \label{eq:Pe_MG_mis}
        \mathrm{Pr}_{\mathrm{e}}\!\!=\!\!\frac{\rho^2}{4\! \sqrt{\pi}}\!\!\sum_{i=1}^{K}\!\!C_i\!\left(\! \frac{2}{a} \!\right)^{\!\frac{\beta_i}{2}}\!\!\!\!{H}^{2,2}_{3,3} \!\left(\!\frac{\sqrt{2} \zeta_i}{a\!\sqrt{\!P_t\!} S_0 |\!h_l\!|}\middle|
        \begin{matrix}
        \!(\!1\!-\!\!\frac{\beta_i}{2},\!1\!)\!,\!(\!\frac{1\!-\beta_i}{2}\!,\!\frac{1}{2}\!),\!(\!1\!\!+\!\rho^2\!\!-\!\beta_i\!,\!1\!)\!\\\!(\rho^2\!-\!\beta_i,\!1),\!
        (0,\!1),\!(\frac{-\beta_i}{2}, \frac{1}{2})\!
        \end{matrix}
        \!\right)\!.
\end{equation}
\end{theorem}

\begin{proposition}[Appendix~\ref{sec:appendix_Pe_no_mis}]
\label{prp:Pe_no_mis}
In the absence of the misalignment effect, the closed-form expression of $\mathrm{Pr}_{\mathrm{e}}$ is
\begin{equation}
    \label{eq:Pe_no_mis}
        \mathrm{Pr}_{\mathrm{e}}\!=\!\frac{1}{2}\!\!\sum_{i=1}^K \!\frac{\alpha_{i}  2^{\frac{\beta_i}{2}}}{|h_l|^{\beta _i} {a}^{\beta_i} \sqrt{\pi} } \!\!\sum_{k=0}^{\infty} \frac{\Gamma\left(\frac{1+\beta_i+ k}{2}\right)}{k !(k+\beta_i)} \left(-\frac{\zeta_i \sqrt{2}}{{a |h_l|}}\right)^k.
\end{equation}
\end{proposition}

\begin{remark}
The $\mathrm{Pr}_{\mathrm{e}}$ expression in \eqref{eq:Pe_no_mis} converges when the parameters of the function $\Phi \left(u,p,r,\omega\right)$ involving the infinite summation in~\eqref{eq:x12} satisfy $r\!<\!2$, $|\text{arg}(\omega)|\!<\! \pi/4$, and the real values of $p$ and $u$ are strictly positive.
\end{remark}

\subsection{Outage probability}
\begin{theorem}[Appendix~\ref{sec:appendix_outage}]
\label{thm:outage}
$\mathrm{Pr}_{\mathrm{out}}$ can be written in a closed form as
\begin{equation}
\label{eq:outage_MG_mis}
        \mathrm{Pr}_{\mathrm{out}}\!=\!\!\sum_{i=1}^{K}\!\!\frac{\rho^2\alpha_i \gamma_{\text{thr}}^{\frac{\beta_i}{2}}}{S_0^{\beta_i} \!|h_l|^{\beta_i} \!\Upsilon^{\frac{\beta_i}{2}}}\!H^{2,1}_{2,3} \!\left(\!\frac{\zeta_i \sqrt{\gamma_{\text {thr }}}}{\sqrt{\Upsilon} S_0 |h_l|} \middle|
        \begin{matrix}
        \!(1\!-\!\beta_i,\!1), (1\!+\!\rho^2\!-\!\beta_i,\!1) \\
        \!(\rho^2\!-\!\beta_i,\!1\!), \!(0,\!1),(-\beta_i,\!1\!)
        \end{matrix}
        \!\right)\!.
\end{equation}
\end{theorem}

\begin{proposition}
In the absence of misalignment, $\mathrm{Pr}_{\mathrm{out}}$ can be expressed in terms of the incomplete gamma function as
\begin{equation}
\label{eq:outage_no_mis}
\begin{aligned}
    \mathrm{Pr}_{\mathrm{out}} & = \frac{1}{2\Upsilon}\sum_{i=1}^K \frac{\alpha_{i}}{\Upsilon^{\frac{\beta _i}{2} -1}|h_l|^{\beta _i }}
    \int_0^{\gamma_{\text {thr }}}x^{\frac{\beta _i}{2} -1} e^{- \frac{\zeta _i}{\sqrt{\Upsilon}|h_l|} \sqrt{x}} d x \\
    & \stackrel{(\mathrm{a})}{=}\frac{1}{\Upsilon}\sum_{i=1}^K \frac{\alpha_{i}}{\Upsilon^{\frac{\beta _i}{2} -1}|h_l|^{\beta _i }}
    \int_0^{\sqrt{\gamma_{\text {thr }}}}t^{\beta _i-1} e^{- \frac{\zeta _i}{\sqrt{\Upsilon}|h_l|} t} d t \\
    & \stackrel{(\mathrm{b})}{=} \sum_{i=1}^K \frac{\alpha_{i}}{\zeta_i^{\beta _i}}\Gamma\left(\beta_i,\frac{\sqrt{\gamma_{\text {thr }}}\zeta _i}{\sqrt{\Upsilon}|h_l|}\right),
\end{aligned}
\end{equation}
where we applied the changes of variables, $t\!=\!\sqrt{x}$ in $\stackrel{(\mathrm{a})}{=}$, $u\!=\! \zeta_it/(\sqrt{\Upsilon}|h_l|)$ in $\stackrel{(\mathrm{b})}{=}$, and used the incomplete gamma function.
\end{proposition}
\subsection{Ergodic capacity}
We next derive the closed-form expressions for $C_{\mathrm{erg}}$.
\begin{theorem}[Appendix~\ref{sec:appendix_capacity}]
\label{thm:capacity}
In the presence of misalignment,
\begin{equation}
\label{eq:capacity_MG_mis}
        C_{\mathrm{erg}}\!=\! \frac{\rho^2}{\ln\!(2\!)}\! \sum_{i=1}^{K}\!\!\frac{\alpha_i}{\zeta_i^{\beta_i}}\!H^{1,4}_{4,3}\!\left(\!\frac{\Upsilon S_0^2 |h_l|^2}{\zeta_i^2} \middle|
        \begin{matrix}
        \!(\!1, \!1),\!(\!1,\!1), \!(\!1\!-\!\!\rho^2, \!2),(\!1\!-\!\beta_i, \!2\!)\! \\
        (1,1),(-\rho^2,2)\!, (0, 1)
        \end{matrix}
        \!\right)\!.
\end{equation}
\end{theorem}

\begin{proposition}
\label{prp:capacity_no_mis}
As in Proposition \ref{prp:Pe_no_mis}, in the absence of misalignment, $C_{\mathrm{erg}}$ is re-expressed using the distribution of $|h| \!=\!|h_l h_f|$,
\begin{equation}
    \label{eq:capacity_MG}
        C_{\mathrm{erg}}\!= \!\!\frac{1}{\ln(2)}\!\!\sum_{i=1}^K \!\!\frac{\alpha_{i}}{|h_l|^{\beta _i }}\!\left(\frac{\zeta _i}{|h_l|}\right)^{\!-\beta_i}\!\!\!H_{3,2}^{1,3} \!\left(\!{\frac{\Upsilon |h_l|^2}{\zeta _i ^2}} \middle|\!\!\begin{array}{c}
        \!(1,\!1\!),\!(1,\!1\!),\!(1\!-\!\!\beta_i,2\!)\!\!\\
        \!(1,\!1),\!(0,1)\!
        \end{array}\!\!\!\right)\!.
\end{equation}
\end{proposition}

\begin{figure*}[t]%
%\vspace{-11mm}
 \centering
 \subfloat[Probability of error]{\label{fig:Pe} \includegraphics[width=0.32\linewidth]{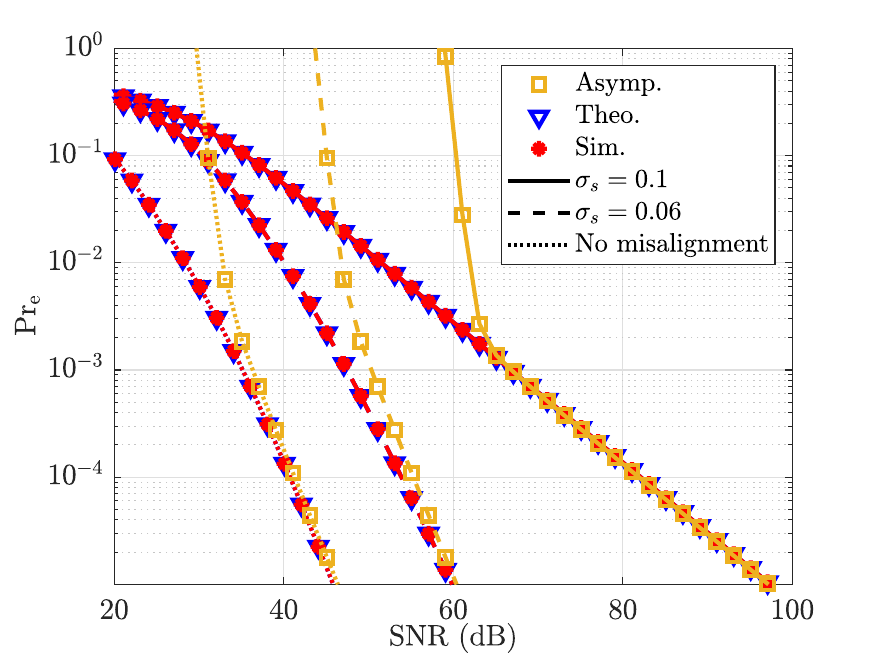}}%
 \hfill
 \subfloat[Outage probability]{\label{fig:outage} \includegraphics[width=0.32\linewidth]{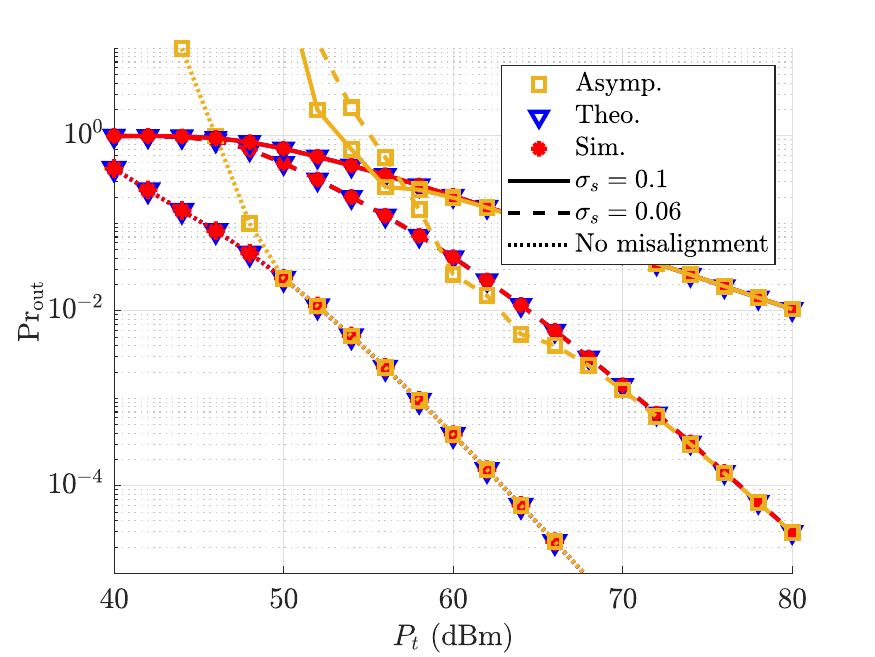}}%
 \hfill
 \subfloat[Ergodic capacity]{\label{fig:capacity} \includegraphics[width=0.32\linewidth]{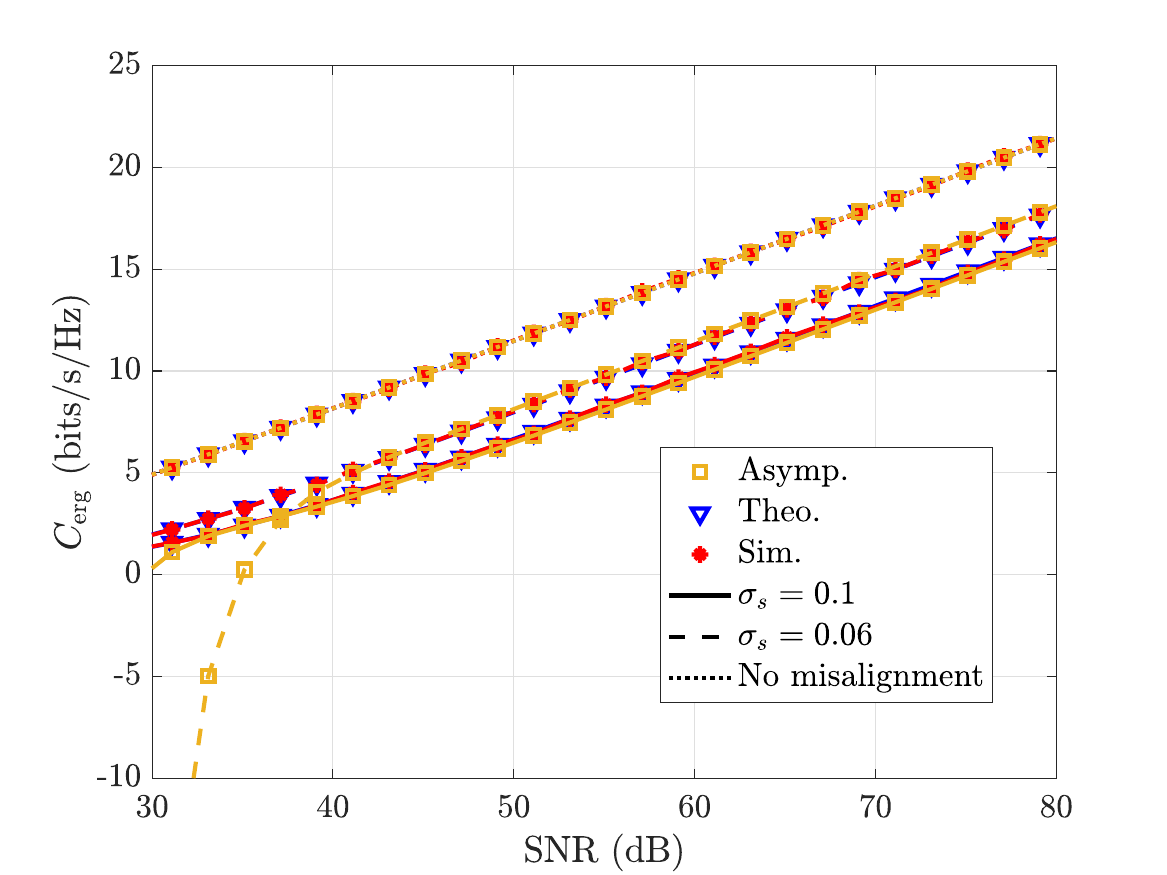}}%
 \caption{Performance evaluation of outdoor THz link under BPSK modulation and MG small-scale fading.}%
 %\vspace{-5mm}
 \label{fig:Capacity}%
\end{figure*}

\section{Asymptotic analysis}

\begin{figure*}[htb]
\begin{align}
    \label{eq:Pe_asymptotic}
    \vspace{-12mm}
        \mathrm{Pr}_{\mathrm{e}} &\underset{z_{\mathrm{P}} \to 0}{\approx} \frac{\rho^2}{4 \sqrt{\pi}} \sum_{i=1}^{K} C_i \left( \frac{2}{a} \right)^{\frac{\beta_i}{2}} \left( \frac{\Gamma\left( \rho^2 \right)\Gamma\left( \frac{1+\rho^2}{2} \right)}{\Gamma\left( 1+\frac{\rho^2}{2} \right)}z_{\mathrm{P}}^{\rho^2 -\beta_i} + \mathcal{O}(z_{\mathrm{P}}^{\rho^2 -\beta_i +1}) + \frac{\Gamma\left( \rho^2 -\beta_i \right)\Gamma\left( \beta_i \right)\Gamma\left( \frac{1+\beta_i}{2} \right)}{\Gamma\left( 1+ \rho^2 -\beta_i \right) \Gamma\left( 1+\frac{\beta_i}{2} \right)} + \mathcal{O}(z_{\mathrm{P}}) \right)\\
        \vspace{-3mm}
        \label{eq:outage_asymptotic}
        \mathrm{Pr}_{\text{out}} &\underset{z_{\mathrm{O}} \to 0}{\approx} \rho^2 \sum_{i=1}^{K} \frac{\alpha_i \gamma_{\text{thr}}^{\frac{\beta_i}{2}}}{S_0^{\beta_i} |h_l|^{\beta_i} \Upsilon^{\frac{\beta_i}{2}}} \left( \frac{\Gamma\left( \rho^2 \right)\Gamma\left( \beta_i - \rho^2 \right)}{\Gamma\left( 1+\rho^2 \right)} z_{\mathrm{O}}^{\rho^2 -\beta_i} + \mathcal{O}(z_{\mathrm{O}}^{\rho^2 -\beta_i +1}) + \frac{\Gamma\left( \rho^2 -\beta_i \right)\Gamma\left( \rho^2 - \beta_i \right)\Gamma\left( \beta_i \right)}{\Gamma\left( 1+ \rho^2 -\beta_i \right) \Gamma\left( 1+\beta_i \right)} + \mathcal{O}(z_{\mathrm{O}}) \right)\\
        \vspace{-3mm}
        \label{eq:foxH_infy}
        H_{p, q}^{m, n}&(z_{\mathrm{C}})\underset{z_{\mathrm{C}} \to \infty}{\approx} {\sum_{i=1}^n}{\vphantom{\sum}}' \left[ h_i z_{\mathrm{C}}^{\frac{a_i-1}{\alpha_i}} + O\left(z^{\frac{a_i-2}{\alpha_i}}\right) \right] + {\sum_{i=1}^n}{\vphantom{\sum}}'' \left( H_i z_{\mathrm{C}}^{\frac{a_i-1}{\alpha_i}} (\log (z_{\mathrm{C}}))^{N_i-1} + O\left(z_{\mathrm{C}}^{\frac{a_i-2}{\alpha_i}} (\log (z_{\mathrm{C}}))^{N_i-1}\right) \right)
\end{align}
\vspace{-6mm}
\end{figure*}

We finally conduct an asymptotic analysis for $\mathrm{Pr}_{\mathrm{e}}$, $\mathrm{Pr}_{\mathrm{out}}$, and $C_{\mathrm{erg}}$ in the high-SNR regime (i.e., $\gamma\!\to\!\infty$). This analysis simplifies the performance metrics from Sec.~\ref{sec:perf_ana}, allowing for easy evaluation of each parameter's effect on the system.

\subsection{Probability of error and outage probability}

\begin{theorem}[Appendix~\ref{sec:appendix_asymptotic_Pe_outage}]
\label{thm:Pe_outage_asymptotic}
As $\gamma \to \infty$, and by setting $z_{\mathrm{P}} = \frac{\sqrt{2} \zeta_i}{a \sqrt{P_t} S_0 |h_l|}$ and $z_{\mathrm{O}} = \frac{\zeta_i \sqrt{\gamma_{\text {thr }}}}{\sqrt{\Upsilon} S_0 |h_l|}$, $\mathrm{Pr}_{\mathrm{e}}$ and $\mathrm{Pr}_{\mathrm{out}}$ are expressed by~\eqref{eq:Pe_asymptotic} and~\eqref{eq:outage_asymptotic}, respectively.
\end{theorem}

\subsection{Ergodic capacity}
Studying the behavior of $C_{\mathrm{erg}}$ in \eqref{eq:capacity_MG_mis} at high SNR requires the asymptotic approximation of the fox-H function when $z_{\mathrm{C}} \!=\! \Upsilon S_0^2 |h_l|^2/\zeta_i^2 \!\to\! \infty$. As illustrated in Appendix~\ref{sec:appendix_asymptotic_Pe_outage}, the asymptotic expansions are valid under constraints \cite{kilbas1999h}. For $C_{\mathrm{erg}}$, the gamma functions have two coinciding poles of indexes 1 and 2 in \eqref{eq:capacity_MG_mis}. Thus, building on \cite[cor. (8)]{kilbas1999h}, the fox-H function expansion when $z_{\mathrm{C}} \!\to\! \infty$ is given by~\eqref{eq:foxH_infy}, where $\sum{\vphantom{\sum}}'$ and $\sum{\vphantom{\sum}}''$ are summations over $i\!\in\!\{1, \cdots, n\}$ \cite[eq. (1.1.1)]{kilbas2004h}, in which the gamma-functions, $\Gamma\left(1-a_i-\alpha_i s\right)$, have simple poles and poles of order $N_i$ in the points $a_{i}$ \cite[eq. (2.24)]{kilbas1999h}, respectively. With $h_i$, $H_i$ defined in \cite[eq. (4.18)]{kilbas1999h}, \cite[eq. (3.14)]{kilbas1999h}, $C_{\mathrm{erg}}$ is expressed at high SNR as
\begin{equation}
        C_{\mathrm{erg}} \!\!\underset{z_{\mathrm{C}} \to\infty}{\approx}\!\! \frac{\rho^2}{\ln(2)}\!\!\sum_{i=1}^{K} \!\!\frac{\alpha_i}{\zeta_i^{\beta_i}}\!\!\left(\!h_3^* z_{\mathrm{C}}^{\frac{-\rho^2}{2}}\!\!+\!h_4^* z_{\mathrm{C}}^{\frac{-\beta_i}{2}}\!\!+\!H_1^* \log(z_{\mathrm{C}}\!)\!+\!H_2^* \log(z_{\mathrm{C}}\!) \!\right)\!.
\end{equation}

\section{Simulation Results and Discussion}

We use the TeraMIMO simulator \cite{tarboush9591285} to accurately model various THz channel characteristics and adapt it to represent small-scale fading using the MG distribution. We define $N_0 = k_{\mathrm{B}}TB$, where $k_{\mathrm{B}}$ the Boltzmann constant, $B$ is the system bandwidth, and $T=\unit[300]{K} $ is the system temperature. Simulations are performed under BPSK modulation with $f=\unit[0.142]{THz}$, $B=\unit[4]{GHz}$, and $d=\unit[64]{m}$. The used antenna gains are $G_r = \unit[19]{dBi}$ and $G_t = \unit[0]{dBi}$ \cite{papasotiriou2023outdoor}. We adjust $P_{t}$ to vary the SNR range. For misalignment fading, we use the same parameters as in \cite{8610080} and vary $\sigma_s$ to change $\rho$.

The theoretical, numerical, and asymptotic bit error probability curves for the outdoor THz channel are illustrated in Fig.~\ref{fig:Pe}. The MG with $K=2$ and relevant parameters are obtained from~\cite{papasotiriou2023outdoor}. Simulations are performed under three configurations: without misalignment and with misalignment ($\sigma_s \in \{0.06, 1\}$). The theoretical plots, derived from equations \eqref{eq:Pe_MG_mis}, \eqref{eq:Pe_no_mis}, and \eqref{eq:Pe_asymptotic}, perfectly match the empirical results, proving the stability of the analytical framework and the validity of the convergence conditions for using Meijer and Fox function properties. A gap of $\unit[15]{dB}$ at $\mathrm{Pr}_{\mathrm{e}}=10^{-4}$ is observed between the no misalignment case and the misalignment case with $\sigma_s = 0.06$. The smaller the value of $\rho$, the more severe the misalignment fading effect, resulting in greater performance degradation. A gap of $\unit[25]{dB}$ at $\mathrm{Pr}_{\mathrm{e}}=10^{-4}$ is noticeable between the configurations $\sigma_s = 0.06$ and $\sigma_s = 0.1$.

The outage probability under MG is illustrated in Fig.~\ref{fig:outage} for $K=2$ and $\gamma_{\mathrm{th}} = 5\, \mathrm{dB}$. Similar to the error probability, there is a match between empirical, theoretical, and asymptotic results in \eqref{eq:outage_MG_mis}, \eqref{eq:outage_no_mis}, and \eqref{eq:outage_asymptotic}. The more severe the misalignment effect, the higher the outage probability at the same transmission power, $P_t$. At $\unit[50]{dBm}$, $\mathrm{Pr}_{\mathrm{out}}$ under no misalignment is slightly below that of the cases of misalignment with $\sigma_s = 0.1$ and $\sigma_s = 0.06$, and around $10^{-2}$ under no misalignment. While the fading for $\sigma_s = 0.06$ is less severe than for $\sigma_s = 0.1$, the outage probability is significantly impacted. Fig.~\ref{fig:capacity} shows the simulated ergodic capacity. Besides an exact match between theoretical and empirical results, there is no difference when varying the number of gamma components between $K=2$ and $K=3$. This observation supports the argument in \cite{papasotiriou2023outdoor} that the number of components, $K$, slightly affects the distribution precision and does not drastically alter performance.

\section{Conclusion}
\label{sec:conclusion}

In this work, we developed a performance analysis framework for outdoor point-to-point SISO THz channels. Leveraging the remarkable fitting traits of the MG distribution in modeling small-scale fading in outdoor THz scenarios and accounting for misalignment fading, we derived tractable performance metrics for bit error probability, outage probability, and ergodic capacity. Furthermore, our theoretical results address the challenges posed by realistic parameters, and the numerical simulations are conducted using state-of-the-art THz measurements. Our theoretical derivations account for the necessary conditions for the validity of the closed-form expressions and are complemented with asymptotic analysis to facilitate tractability at high SNR regimes.

\appendices
\counterwithin*{equation}{section}
\renewcommand\theequation{\thesection.\arabic{equation}}
\section{}
\label{sec:appendix_pdf}
Proof of Theorem \ref{thm:pdfh}: Using \eqref{MGdist}, we start by deriving the PDF of $ |h_{f} h_{m}|$ as
\begin{equation}
\begin{aligned}
&f_{|h_{f} h_{m}|}(x)\!=\!\!\int_0^{S_0}\!\!\frac{1}{y} f_{h_{f}}\!\left(\frac{x}{y}\right) f_{h_m}(y) dy\!=\!\! \frac{{\rho ^2}}{S_0^{{\rho ^2}}} \!\!\sum_{i=1}^{K}\!\!\int_0^{S_0}\!\!\! y^{{\rho ^2}\!-\!\beta_i} \!e^{-\frac{\zeta_i x}{y}}\!dy\\
&\stackrel{(\mathrm{a})}{=} {\rho ^2 } \sum_{i=1}^{K} \frac{\alpha_i x^{\beta_i-1}}{S_0^{\beta_i }} \int_0^{1} u^{{\rho ^2} - \beta_i -1} e^{-\frac{\zeta_i x}{S_0 u}} \, du \\
&\stackrel{(\mathrm{b})}{=} {\rho ^2} \sum_{i=1}^{K} \frac{\alpha_i x^{\beta_i-1}}{S_0^{\beta_i }} \int_1^{\infty} t^{\beta_i - {\rho ^2} - 1} e^{-\frac{\zeta_i x}{S_0} t} \, dt \\
&\stackrel{(\mathrm{c})}{=} \sum_{i=1}^{K} \frac{\alpha_i x^{\beta_i-1}}{\rho ^{-2} S_0^{\beta_i }} \int_1^{\infty} t^{-(1+ {\rho ^2}-\beta_i)} G_{0,1}^{1,0} \left[ \frac{\zeta_i x}{S_0} t \Bigg| \begin{array}{c} \sim \\ 0 \end{array} \right]  \, dt \\
&\stackrel{(\mathrm{d})}{=} {\rho ^2} \sum_{i=1}^{K} \frac{\alpha_i x^{\beta_i-1}}{S_0^{\beta_i }} G_{1,2}^{2,0} \left( \left. \frac{\zeta_i x}{S_0} \, \right| 
\begin{array}{l}
1 + {\rho ^2} - \beta_i \\
{\rho ^2} - \beta_i, 0
\end{array} \right),
\end{aligned}
\end{equation}
where we apply the changes of variables $u = y/S_0$ and $t = 1/u$ in $\stackrel{(\mathrm{a})}{=}$ and $\stackrel{(\mathrm{b})}{=}$, respectively. In $\stackrel{(\mathrm{c})}{=}$, we express $e^{-\frac{\zeta_i x}{S_0} t}$ in terms of the Meijer function as $e^{-\frac{\zeta_i x}{S_0} t} =G_{0,1}^{1,0} \left[ \frac{\zeta_i x}{S_0} t \Bigg| \begin{array}{c} \sim \\ 0 \end{array} \right]$. Finally, we use the Euler property of the Meijer function in $\stackrel{(\mathrm{d})}{=}$ \cite[eq. (7.811.3)]{prudnikov1986integrals}. The PDF of $|h|$ is obtained by using $f_{|h|}(x)=\frac{1}{|h_l|}f_{|h_f h_m|}\left(\frac{x}{|h_l|}\right)$ which results in \eqref{thm:pdf_h}. To get \eqref{thm:pdf_h_squared_Pt}, we use the transformation $Y=X^2=g(X)$ following the property $f_Y(y)=f_X\left(g^{-1}(y)\right) |\frac{d}{dy}g^{-1}(y)|$.
\section{}
\label{sec:appendix_Pe}
Proof of Theorem \ref{thm:Pe}: By replacing \eqref{thm:pdf_h_squared_Pt} in \eqref{eq:prob_error}, we get
\begin{equation}
\begin{aligned}
    &\mathrm{Pr}_{\mathrm{e}}\!=\!\!\int_0^{\infty}\!\!\!{Q}(a\sqrt{x}) \frac{\rho^2}{2} \!\!\sum_{i=1}^{K}\!C_i x^{\frac{\beta_i}{2}-1} \!G^{2,0}_{1,2}\left(\! \frac{\zeta_i \sqrt{x}}{\sqrt{P_t} S_0 |h_l|} \middle| 
    \begin{matrix}
    1 \!+\! \rho^2 \!-\! \beta_i \\ 
    \rho^2 \!-\! \beta_i , 0
    \end{matrix}
    \right) \,\!dx \\
    &\stackrel{(\mathrm{a})}{=}\! \frac{\rho^2}{2}\!\sum_{i=1}^{K} C_i \!\!\int_0^{\infty}\!\!\!{Q}(a\sqrt{x}) x^{\frac{\beta_i}{2} - 1} G^{2,0}_{1,2} \left(\! \frac{\zeta_i \sqrt{x}}{\sqrt{P_t} S_0 |h_l|} \middle| 
    \begin{matrix}
    1 \!+\! \rho^2 \!-\! \beta_i \\ 
    \rho^2 \!-\! \beta_i , 0
    \end{matrix}
    \right) \, dx \\
    &\!\stackrel{(\mathrm{b})}{=}\!\!
    \frac{\rho^2}{4\!\sqrt{\pi}}\!\!\sum_{i=1}^{K}\!\!C_i\!\!\!\int_0^{\infty}\!\!\!\!x^{\frac{\beta_i}{2}\!-\!1}\!G^{2,0}_{1,2}\!\left(\! \frac{ax}{2} \middle|
    \begin{matrix}
    1 \\
    0,\!\frac{1}{2}
    \end{matrix}
    \!\right)\!G^{2,0}_{1,2} \!\left( \!\frac{\zeta_i \sqrt{x}}{\sqrt{P_t} S_0 |h_l|} \middle|
    \begin{matrix}
    1 \!+\! \rho^2\!\!-\!\beta_i \\
    \rho^2\!-\!\beta_i , \!0
    \end{matrix}
    \!\right)\!dx,
\end{aligned}
\end{equation}
where in $\stackrel{(\mathrm{a})}{=}$, we replace the $Q$ function by \({Q}(a\sqrt{x}) = \frac{1}{2} \text{erfc} \left( \frac{\sqrt{a^2 x}}{2} \right)\). In $\stackrel{(\mathrm{b})}{=}$, we write the erfc function in its Meijer form using \(\text{erfc}(\sqrt{x}) = \frac{1}{\sqrt{\pi}} {G}^{2,0}_{1,2} \left( x \middle|
\begin{matrix}
1 \\
0, \frac{1}{2}
\end{matrix}
\right)\). Finally, we use the Meijer integration property \cite{website_wolfram} to obtain~\eqref{eq:Pe_MG_mis}.
\section{}
\label{sec:appendix_Pe_no_mis}
Proof of Proposition \ref{prp:Pe_no_mis}: We replace the distribution of $|h|^2 = |h_l h_f|^2$ in \eqref{eq:prob_error} to obtain
\begin{equation}
\begin{aligned}
\label{eq:x11}
    \mathrm{Pr}_{\mathrm{e}} & =\frac{1}{2}\sum_{i=1}^K \frac{\alpha_{i}}{|h_l|^{\beta_i }}\int_0^{+\infty} Q(a\sqrt{x} ) x^{\frac{\beta _i}{2} -1} e^{- \frac{\zeta _i}{|h_l|} \sqrt{x}} d x \\
    &  \stackrel{(\mathrm{a})}{=}\sum_{i=1}^K \frac{\alpha_{i}}{|h_l|^{\beta _i }}\int_0^{+\infty} Q(at ) t^{\beta_i -1} e^{- \frac{\zeta _i}{|h_l|} t} d t \\
    & \stackrel{(\mathrm{b})}{=}\frac{1}{2}\sum_{i=1}^K \frac{\alpha_{i}}{|h_l|^{\beta _i }}\int_0^{+\infty} \mathrm{erfc}(\frac{at}{\sqrt{2}}) t^{\beta_i -1} e^{- \frac{\zeta _i}{|h_l|} t} d t \\
    & \stackrel{(\mathrm{c})}{=}\frac{1}{2}\sum_{i=1}^K \frac{\alpha_{i}}{|h_l|^{\beta _i }}\Phi\left(\beta_i, \frac{\zeta_i}{|h_l|}, 1, \frac{a}{\sqrt{2}}\right),
\end{aligned}
\end{equation}
where we apply the variable change $t = \sqrt{x}$ in $\stackrel{(\mathrm{a})}{=}$, and utilize ${Q}(x) = \frac{1}{2} \mathrm{erfc}\left(\frac{x}{\sqrt{2}}\right)$ in $\stackrel{(\mathrm{b})}{=}$. To obtain \eqref{eq:Pe_no_mis}, we use in $\stackrel{(\mathrm{c})}{=}$ the $\Phi$ function defined as \cite[eq. (2.8.1.5)] {prudnikov1986integrals} \begin{equation} 
        \label{eq:x12}
        \Phi\left(u,p,r,\omega\right)  \!=\!\frac{1}{\omega^{u} \sqrt{\pi}} \!\!\sum_{k=0}^{\infty}\!\frac{1}{k !(r k\!+\!u)}\! \Gamma\left(\!\frac{1\!+\!u\!+\!r k}{2}\!\right)\!\left(\!\frac{-p}{{\omega}^r}\right)^k\!.
\end{equation}
\section{}
\label{sec:appendix_outage}
Proof of Theorem \ref{thm:outage}: Replacing the PDF of $\gamma\!=\!\Upsilon|h|^2$ in \eqref{eq:out},
\begin{equation}
\begin{aligned}
    &\mathrm{Pr}_{\mathrm{out}}\!=\! \frac{\rho^2}{2}\!\!\sum_{i=1}^{K} \!\!\int_0^{\gamma_{\text{thr}}}\!\!\! \frac{\alpha_i x^{\frac{\beta_i}{2} - 1}}{S_0^{\beta_i} |h_l|^{\beta_i} \Upsilon^{\frac{\beta_i}{2}}}
     {G}^{2,0}_{1,2} \left( \frac{\zeta_i \sqrt{x}}{\sqrt{\Upsilon} S_0 |h_l|} \middle|
    \begin{matrix}
    1 \!+\! \rho^2 \!-\! \beta_i \\
    \rho^2 \!-\! \beta_i, 0
    \end{matrix}
    \right)\!dx\\
    &\stackrel{(\mathrm{a})}{=}\!\frac{\rho^2}{2}\!\!\sum_{i=1}^{K} \int_0^{1}\!\!\!\frac{\alpha_i \gamma_{\text {thr }}^{\frac{\beta_i}{2}} u^{\frac{\beta_i}{2} - 1}}{S_0^{\beta_i} |h_l|^{\beta_i} \Upsilon^{\frac{\beta_i}{2}}}  {G}^{2,0}_{1,2} \left( \frac{\zeta_i \sqrt{\gamma_{\text {thr }}} \sqrt{u}}{\sqrt{\Upsilon} S_0 |h_l|} \middle|
    \begin{matrix}
    1 + \rho^2 - \beta_i \\
    \rho^2 - \beta_i, 0
    \end{matrix}
    \right) du \\
    &\stackrel{(\mathrm{b})}{=}\!\rho^2\!\! \sum_{i=1}^{K} \int_0^{1}\!\!\!\frac{\alpha_i \gamma_{\text {thr }}^{\frac{\beta_i}{2}} t^{\beta_i - 1}}{S_0^{\beta_i} |h_l|^{\beta_i} \Upsilon^{\frac{\beta_i}{2}}}  {G}^{2,0}_{1,2} \left( \frac{\zeta_i t}{\sqrt{\Upsilon} S_0 |h_l|} \middle|
    \begin{matrix}
    1 + \rho^2 - \beta_i \\
    \rho^2 - \beta_i, 0
    \end{matrix}
    \right) dt \\
    &\stackrel{(\mathrm{c})}{=}\!\rho^2\!\!\sum_{i=1}^{K}\!\frac{\alpha_i \gamma_{\text {thr }}^{\frac{\beta_i}{2}}}{S_0^{\beta_i} |h_l|^{\beta_i} \Upsilon^{\frac{\beta_i}{2}}} {G}^{2,1}_{2,3} \left( \frac{\zeta_i \sqrt{\gamma_{\text {thr }}}}{\sqrt{\Upsilon} S_0 |h_l|} \middle|
    \begin{matrix}
    1 - \beta_i, 1 + \rho^2 - \beta_i \\
    \rho^2 - \beta_i, 0,-\beta_i
    \end{matrix}
    \right),
\end{aligned}
\end{equation}
where we apply the variables changes, $u \!=\! x/\gamma_{\mathrm{thr}}$ in $\stackrel{(\mathrm{a})}{=}$, and $t\!=\!\sqrt{u}$ in $\stackrel{(\mathrm{b})}{=}$, and use the Euler property of the Meijer G-function \cite[eq. (7.811.2)]{prudnikov1986integrals}. Finally, replacing the Meijer G-function with its fox-H equivalent~\cite{kilbas2004h} results in $\mathrm{Pr}_{\text {out }}$ in \eqref{eq:outage_MG_mis}.
\section{}
\label{sec:appendix_capacity}
Proof of Theorem \ref{thm:capacity}: For ergodic capacity, we substitute $|h|$ in \eqref{eq:capacity}, resulting in
\begin{equation}
    \begin{aligned}
        &C_{\mathrm{erg}}\!\!=\!\!\frac{1}{\ln\!(2\!)}\!\!\int_0^{\infty}\!\!\!\ln(\!1\!\!+\!\!\Upsilon x^2) \rho^2\!\!\sum_{i=1}^{K} \!\!\frac{\alpha_i x^{\beta_i-1}}{S_0^{\beta_i} \!|h_l|^{\beta_i}} \!G_{1,2}^{2,0}\!\left(\! \frac{\zeta_i x}{S_0 |h_l|}\middle|\!\! 
    \begin{array}{c}
    \!1\!+\!\!\rho^2\!-\!\beta_i \\
    \rho^2\!\!-\!\beta_i,\!0
    \!\end{array}\!\!\!\!\right)\!dx \\
    &\!=\!\frac{\rho^2}{\ln\!(2\!)}\!\! \sum_{i=1}^{K}\!\!\frac{\alpha_i}{S_0^{\beta_i}\! |h_l|^{\beta_i}}\!\!\int_0^{\infty}\!\!\!x^{\beta_i-1}\!\ln(\!1\!\!+\!\!\Upsilon x^2) G^{2,0}_{1,2}\!\left( \!\frac{\zeta_i x}{S_0 |h_l|} \middle| 
    \begin{matrix}
    1\!+\!\rho^2\!\!-\!\beta_i\!\\
    \rho^2\!\!-\!\beta_i,\!0\!
    \end{matrix}
    \!\right)\!dx \\
    &\!\stackrel{(\mathrm{a})}{=}\!\!\!\frac{\rho^2}{\ln\!(2\!)}\!\! \sum_{i=1}^{K}\!\!\frac{\alpha_i}{S_0^{\beta_i} \!|h_l|^{\beta_i}} \!\!\int_0^{\infty}\!\!\!\!\!x^{\beta_i\!- 1}\! G^{1,2}_{2,2}\!\left(\! \Upsilon x^2\middle|
    \begin{matrix}
    1\!,\!1 \\
    1\!,\!0
    \end{matrix}
    \!\right)\!\!G^{2,0}_{1,2}\!\!\left(\! \frac{\zeta_i x}{S_0 \!|h_l|} \middle|
    \begin{matrix}
    \!1\!+\!\rho^2\!\!-\!\beta_i\!\\
    \!\rho^2\!\!-\!\beta_i,\!0\!
    \end{matrix}
    \!\right)\!dx 
\end{aligned}
\end{equation}
where in $\stackrel{(\mathrm{a})}{=}$, we replace the logarithm function by $\ln (1+\Upsilon x^2) = G_{2,2}^{1,2} \left[ \Upsilon x^2 \Bigg| \begin{array}{c} 1,1 \\ 1,0 \end{array} \right]$, and the final result in \eqref{eq:capacity_MG_mis} is obtained using the Meijer G-function integration property \cite{website_wolfram}.

\section{}
\label{sec:appendix_asymptotic_Pe_outage}
Proof of Theorem \ref{thm:Pe_outage_asymptotic}: For $\mathrm{Pr}_{\mathrm{e}}$ and $\mathrm{Pr}_{\mathrm{out}}$, we use the asymptotic expansion of the fox H-function near zero~\cite{kilbas1999h},
\begin{equation}
\label{eq:fox_near_0}
H_{p, q}^{m, n}(z) = \sum_{j=1}^m \left[ h_j^* z^{b_j / \beta_j} + O\left(z^{(b_j+1) / \beta_j}\right) \right] \quad (z \rightarrow 0),
\end{equation}
\begin{equation}
h_j^* \! =\!\frac{\prod_{i=1, i \neq j}^p \Gamma\left(b_i - \frac{b_j \beta_i}{\beta_j}\right) \prod_{i=1}^n \Gamma\left(1 - a_i + \frac{b_j \alpha_i}{\beta_j}\right)}{\beta_j\prod_{i=n+1}^p \Gamma\left(a_i - \frac{b_j \alpha_i}{\beta_j}\right) \prod_{i=m+1}^q \Gamma\left(1 - b_i + \frac{b_j \beta_i}{\beta_j}\right)}\!.\!
\end{equation}
This expansion is valid when $\Delta = \sum \beta_i - \sum \alpha_i \geq 0$, and under conditions \cite[cond. (1.6)]{kilbas1999h} and \cite[cond. (1.7)]{kilbas1999h}, validated in our system model for $\mathrm{Pr}_{\mathrm{e}}$ and $\mathrm{Pr}_{\mathrm{out}}$. By replacing the fox-H function with its asymptotic expression in \eqref{eq:Pe_MG_mis} and \eqref{eq:outage_MG_mis},
\begin{equation}
        \mathrm{Pr}_{\mathrm{e}}\underset{z_{\mathrm{P}}\to 0}{\approx}\!\frac{\rho^2}{4 \sqrt{\pi}}\!\sum_{i=1}^{K}\!\! C_i\!\left( \frac{2}{a} \right)^{\frac{\beta_i}{2}} \!\!\left(\!h_1^{*} z_{\mathrm{P}}^{\rho^2 -\beta_i}\!+\!\mathcal{O}(z_{\mathrm{P}}^{\rho^2 -\beta_i +1})\!+\!h_2^{*}\!+\! \mathcal{O}(z_{\mathrm{P}}) \right),
\end{equation}

where $z_{\mathrm{P}} = \frac{\sqrt{2} \zeta_i}{a \sqrt{P_t} S_0 |h_l|}$. Replacing $h_1^*$ and $h_2^*$ with their expressions leads to the final result in \eqref{eq:Pe_asymptotic}. Similar procedure with $z_{\mathrm{O}} = \frac{\zeta_i \sqrt{\gamma_{\text {thr }}}}{\sqrt{\Upsilon} S_0 |h_l|}$ leads to the asymptotic expression in \eqref{eq:outage_asymptotic}.

\bibliographystyle{ieeetr}

\begin{thebibliography}{99}

\bibitem{sarieddeen2020overview}
H. Sarieddeen, M.-S. Alouini, and T. Y. Al-Naffouri, ``An overview of signal processing techniques for terahertz communications,'' \textit{Proc. IEEE}, vol. 109, no. 10, pp. 1628--1665, 2021.

\bibitem{Jornet2024Evolution}
J. M. Jornet \textit{et al.}, ``The evolution of applications, hardware design, and channel modeling for terahertz (THz) band communications and sensing: Ready for 6G?,'' \textit{Proc. IEEE}, pp. 1--32, 2024.

\bibitem{sheikh2022thz}
F. Sheikh \textit{et al.}, ``THz measurements, antennas, and simulations: From the past to the future,'' \textit{IEEE Journal of Microwaves}, vol. 3, no. 1, pp. 289--304, 2022.

\bibitem{tarboush9591285}
S. Tarboush \textit{et al.}, ``TeraMIMO: A channel simulator for wideband ultra-massive MIMO terahertz communications,'' \textit{IEEE Trans. on Vehic. Technol.}, vol. 70, no. 12, pp. 12 325--12 341, 2021.

\bibitem{papasotiriou2023outdoor}
E. N. Papasotiriou, A.-A. A. Boulogeorgos, and A. Alexiou, ``Outdoor THz fading modeling by means of Gaussian and gamma mixture distributions,'' \textit{Sci. Rep.}, vol. 13, no. 1, p. 6385, 2023.

\bibitem{karakoca2023measurement}
E. Karakoca \textit{et al.}, ``Measurement-based modeling of short range terahertz channels and their capacity analysis,'' in \textit{Proc. IEEE Global Commun. Conf. (GLOBECOM)}, 2023, pp. 1471--1476.

\bibitem{papasotiriou2021experimentally}
E. N. Papasotiriou \textit{et al.}, ``An experimentally validated fading model for THz wireless systems,'' \textit{Sci. Rep.}, vol. 11, no. 1, p. 18717, 2021.

\bibitem{atapattu2011mixture}
S. Atapattu, C. Tellambura, and H. Jiang, ``A mixture gamma distribution to model the SNR of wireless channels,'' \textit{IEEE Trans. Wireless Commun.}, vol. 10, no. 12, pp. 4193--4203, 2011.

\bibitem{jung2014capacity}
J. Jung \textit{et al.}, ``Capacity and error probability analysis of diversity reception schemes over generalized-K fading channels using a mixture gamma distribution,'' \textit{IEEE Trans. Wireless Commun.}, vol. 13, no. 9, pp. 4721--4730, 2014.

\bibitem{8610080}
A.-A. A. Boulogeorgos, E. N. Papasotiriou, and A. Alexiou, ``Analytical performance assessment of THz wireless systems,'' \textit{IEEE Access}, vol. 7, pp. 11 436--11 453, 2019.

\bibitem{10018285}
O. S. Badarneh, M. T. Dabiri, and M. Hasna, ``Channel modeling and performance analysis of directional THz links under pointing errors and $\alpha-\mu$ distribution,'' \textit{IEEE Commun. Lett.}, vol. 27, no. 3, pp. 812--816, 2023.


\bibitem{9714471}
O. S. Badarneh, ``Performance analysis of terahertz communications in random fog conditions with misalignment,'' \textit{IEEE Wireless Commun. Lett.}, vol. 11, no. 5, pp. 962--966, 2022.

\bibitem{bhardwaj2022performance}
P. Bhardwaj and S. M. Zafaruddin, ``Performance analysis of outdoor THz wireless transmission over mixed Gaussian fading with pointing errors,'' in \textit{Proc. Int. Conf. on Next Gen. Syst. and Networks (ICNGSN)}, 2022, pp. 187--196.

\bibitem{Varotsos2023capacity}
G. K. Varotsos \textit{et al.}, ``Capacity performance analysis for terrestrial THz wireless channels,'' \textit{Electronics}, vol. 12, no. 6, p. 1336, 2023.

\bibitem{10458985}
Y. Li and Y. J. Chun, ``Analysis of IRS-assisted downlink wireless networks over generalized fading,'' \textit{IEEE Trans. Wireless Commun.}, 2024, early access.

\bibitem{prudnikov1986integrals}
A. P. Prudnikov \textit{et al.}, \textit{Integrals and series: Special functions}, CRC press, 1986, vol. 2.

\bibitem{kilbas2004h}
A. A. Kilbas, \textit{H-transforms: Theory and Applications}, CRC press, 2004.

\bibitem{tarboush2022single}
S. Tarboush \textit{et al.}, ``Single- versus multicarrier terahertz-band communications: A comparative study,'' \textit{IEEE Open J. of the Commun. Soc.}, vol. 3, pp. 1466--1486, August 2022.

\bibitem{simon2001digital}
M. K. Simon and M.-S. Alouini, \textit{Digital communication over fading channels}. John Wiley \& Sons, 2005, vol. 95.

\bibitem{kilbas1999h}
A. A. Kilbas and M. Saigo, ``On the H-function,'' \textit{Int. Journal of Stochastic Analysis}, vol. 12, no. 2, pp. 191--204, 1999.

\bibitem{website_wolfram}
Meijer G-function: integration. [Online]. Available: \url{https://functions.wolfram.com/07.34.21.0012.01}

\end{thebibliography}

\end{document}